\documentclass[twocolumn]{aastex631}

\newcommand{\source}{MAXI~J1752$-$457}
\newcommand{\fluxcgs}{\ensuremath{\mathrm{erg\,cm^{-2}\,s^{-1}}}}
\newcommand{\lumcgs}{\ensuremath{\mathrm{erg\,s^{-1}}}}

\usepackage{multirow}

\begin{document}

\title{Investigating \source\ with NuSTAR in the aftermath of a superburst}

\correspondingauthor{Sean N. Pike}
\email{seannpike.astro@gmail.com}

\author[0000-0002-8403-0041]{Sean N. Pike}
\affiliation{Department of Astronomy \& Astrophysics, University of California, San Diego, 9500 Gilman Dr, La Jolla, CA 92093, USA}

\author[0000-0003-0939-1178]{Hitoshi Negoro}
\affiliation{Department of Physics, Nihon University, 1-8 Kanda-Surugadai, Chiyoda-ku, Tokyo 101-8308, Japan}

\author[0000-0002-5341-6929]{Douglas Buisson}
\affiliation{Independent Researcher}

\author[0000-0003-0870-6465]{Benjamin Coughenour}
\affiliation{Department of Physics, Utah Valley University, 800 W. University Pkwy., MS 179, Orem, UT 84508, USA}

\author[0000-0001-6470-6553]{Julian Gerber}
\affiliation{Department of Astronomy \& Astrophysics, University of California, San Diego, 9500 Gilman Dr, La Jolla, CA 92093, USA}

\author[0000-0002-8808-520X]{Aarran W. Shaw}
\affiliation{Department of Physics \& Astronomy, Butler University, 4600 Sunset Avenue, Indianapolis, IN 46208, USA}

\author[0000-0002-1190-0720]{Mutsumi Sugizaki}
\affiliation{Advanced Research Center for Space Science and Technology, Kanazawa University, Kakuma, Kanazawa, Ishikawa 920-1192, Japan}

\author[0000-0001-5506-9855]{John A. Tomsick}
\affiliation{Space Sciences Laboratory, 7 Gauss Way, University of California, Berkeley, CA 94720-7450, USA}








\begin{abstract}

We present two NuSTAR observations of the X-ray transient, \source, following a superburst which was observed by MAXI/GSC in November, 2024. NuSTAR follow-up confirmed that \source\ is coincident with the previously observed Einstein Probe source, EP240809a. We performed a spectral analysis of the source during both NuSTAR observations, and we find that the hard X-ray spectra are consistent with the inclusion of a spherical blackbody component and a steep, non-thermal, power law component. At about 79\,hours after the onset of the superburst, we find a blackbody temperature of $kT_\mathrm{bb}=0.60\pm0.1$\,keV and $R_\mathrm{bb}/D_{8}=6.0^{+0.4}_{-0.3}$\,km (not including corrections for scattering in the neutron star atmosphere), where $D_{8}$ is the source distance, which is not yet known, in units of 8\,kpc. We found that the blackbody temperature did not change significantly in the one day interval between successive NuSTAR observations, and we performed an energy-resolved timing analysis which showed that the source variability was dominated by red noise in the power law component, suggesting coupling with an accretion disk. We infer that the source had entered an accretion-powered flux state. Furthermore, we measure a photon index of $\Gamma\approx4$, much steeper than those typically observed during accretion onto neutron stars at similar luminosities. This is suggestive of ongoing evolution of the electron energy distribution responsible for the power law component several days after the superburst, but the lack of hard X-ray observations prior to and throughout the superburst make it difficult to present a conclusive physical interpretation of this result.
\end{abstract}


\section{Introduction} \label{sec:intro}

X-ray binaries (XRB), formed via accretion onto a compact object from a companion, exhibit a wide array of variability on timescales ranging from milliseconds to days to months. In particular, most XRBs are transient, meaning that they alternate between intervals of quiescence and outburst due to changes in the rate of mass accretion. For example, in Be X-ray binaries consisting of a Be star and an accreting neutron star, these periods of outburst are caused by transient accretion when the compact object passes near its companion in its orbit. XRBs hosting black holes undergo weeks- or months-long outbursts due to slow accumulation of material into their accretion disks, leading to a critical point at which material accretes much more efficiently onto the compact object \citep{DIM}.

These accretion-powered outbursts are distinct from bursting behavior caused by thermonuclear burning. Neutron star XRBs which persistently accrete from low mass companions at a relatively low level may accumulate material on their surfaces without burning until a critical depth is reached, producing runaway burning of accreted hydrogen and helium. We observe this runaway conflagration in the form of a rapid flash known as a Type I X-ray burst, lasting only seconds to minutes, during which the source luminosity may increase by multiple orders of magnitude up to the Eddington luminosity. The depth at which accumulated material ignites plays a decisive role in the duration and structure of the resulting burst. Some sources exhibit ``superbursts" which resemble Type I X-ray bursts but which last several hours and decay on the timescale of days due to ignition at much larger column depths producing extended carbon burning \citep{Zand2017,Cumming2003,Keek2008,Serino2016}. Due to their extended timescale, superbursts can provide an opportunity to probe how accretion disks and coronae respond to high levels of X-ray irradiation. The rarity of these events, however, has thus far limited investigations into superbursts. 
In this paper, we present an analysis of the newly-discovered source, \source, which was observed by the Nuclear Spectroscopic Telescope Array (NuSTAR) during the decay phase of a superburst.

The first detection of \source\ with the Monitor of All-sky X-ray Image (MAXI) Gas Slit Camera (GSC) occurred during the scan transit at 18:23 on November 9, 2024 UTC (MJD 60623.766) with a 4--10\,keV flux of $982\pm50$\,mCrab, or $(1.12\pm0.06)\times10^{-8}$\,\fluxcgs \citep{Serino2024}. The flux dropped rapidly over subsequent scan transits. Several months prior to the detection of \source\ with MAXI, an X-ray transient, deemed EP240809a, was detected by Einstein Probe (EP) with a position consistent with \source\ \citep{Liu2024}. This earlier period of brightening, which was observed on August 9, 2024 (UTC), reached a much lower flux than the November burst: \cite{Liu2024} reported an unabsorbed 0.4--5\,keV flux of $5.3\times10^{-11}$\fluxcgs, corresponding to about 1.5\,mCrab. An optical flare from the source was also observed by the Asteroid Terrestrial-impact Last Alert System (ATLAS) beginning about three weeks before the EP detection \citep{Zhu2024}. The authors pointed out that the soft X-ray and the blue optical spectra were consistent with a tidal disruption event (TDE), 
but no host galaxy was found. Therefore, the nature of the source remained unknown.

MAXI/GSC observations showed a thermal spectrum which softened during the rapid decay, leading \cite{Negoro2024} to propose that the origin of the source activity was a long X-ray burst from an uncatalogued X-ray binary, rather than a TDE. By combining MAXI/GSC and NinjaSat observations of the November burst, \cite{Aoyama2025} have shown that the source is indeed consistent with an accreting neutron star. The flux evolution of the burst, which exhibits a fast rise and exponential decay on the timescale of days, along with an X-ray spectrum which can be described with a spherical blackbody, indicate that the transient event corresponded to accretion-induced thermonuclear burning. In particular, the hours-long timescale of the burst and the years-long recurrence time (inferred from the lack of previous detections of the source by MAXI), indicate a superburst.

At the time that it was detected by MAXI/GSC, \source\ had a small Solar Aspect Angle (SAA) of $<45^\circ$, making follow-up observations impossible for most pointing X-ray observatories. Thanks to the ability of NuSTAR to point near or at the Sun, we were able to trigger hard X-ray observations of the MAXI confidence region. These observations confirmed that \source\ is coincident with EP240809a, and preliminary spectral analysis provided the first evidence for a hard X-ray excess in addition to the blackbody spectral component \citep{Pike2024}.

Here, we present a detailed analysis of two NuSTAR observations of \source\ following its November, 2024, superburst, and we place the flux of the source as measured by NuSTAR in the context of MAXI/GSC monitoring of the burst. In Section \ref{sec:maxi}, we briefly present the MAXI/GSC observations of the burst. In Section \ref{sec:nustar}, we present both spectral and timing analyses of the source with NuSTAR. In Section \ref{sec:discussion}, we discuss the implications of these results, and finally we list our conclusions in Section \ref{sec:conclusions}. All errors are quoted at the $90\%$ confidence level unless otherwise stated.

\section{MAXI light-curve} \label{sec:maxi}

\begin{deluxetable*}{cccc}
\tablenum{1}
\tablecaption{Details of the NuSTAR observations. \label{tab:nustar_obs}}
\tablewidth{0pt}
\tablehead{
\colhead{OBSID} &  \colhead{Start time (MJD)}    & \colhead{End time (MJD)}   & \colhead{Exposure (ks)}}
\startdata
81010302001 &   60626.32    &   60626.40    &   6.6 \\
81001347001 &   60627.05    &   60627.55    &   26
\enddata
\end{deluxetable*}

\begin{figure*}[t!]
\begin{center}
\includegraphics[width=\textwidth]{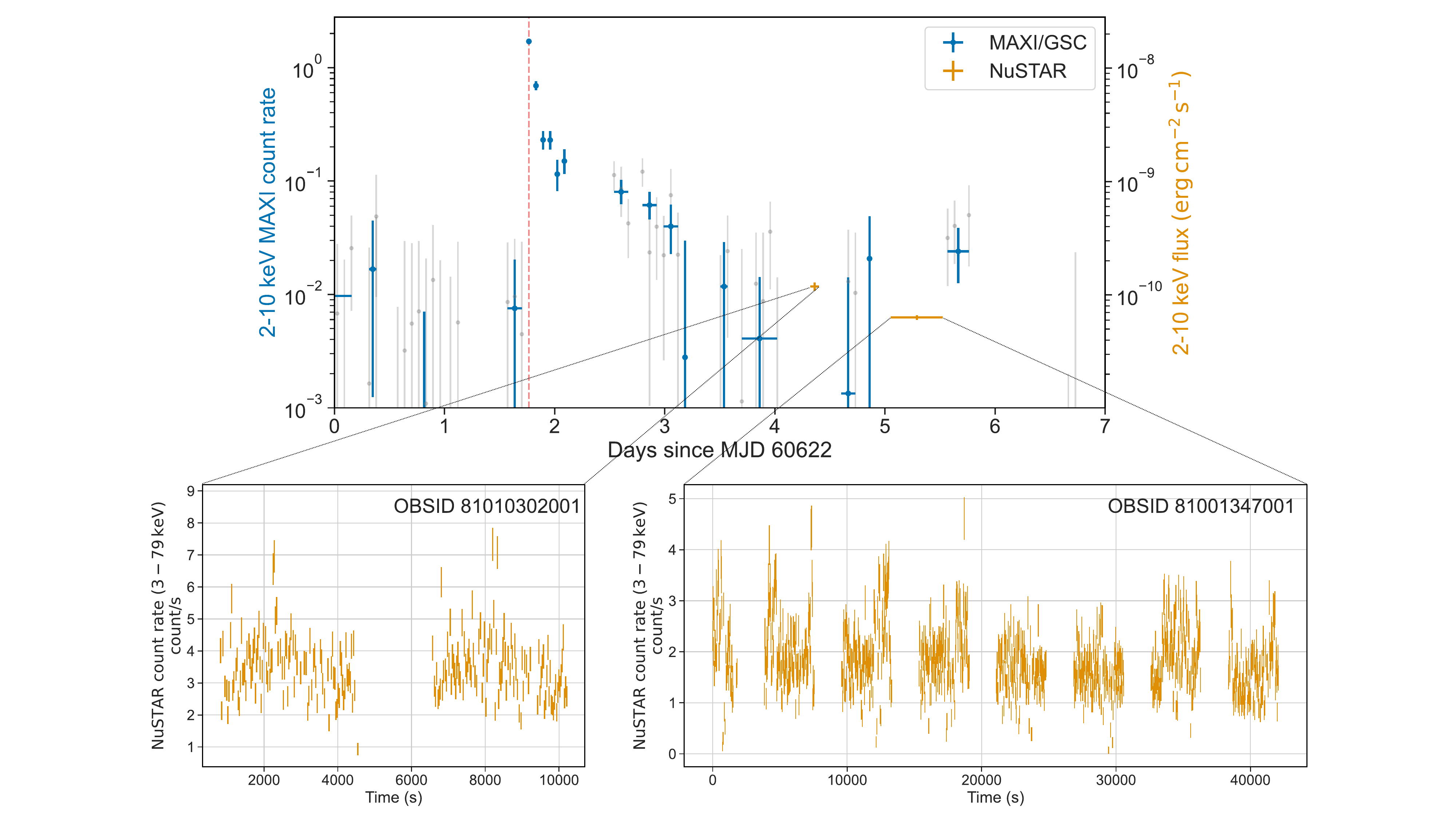}
\caption{Top: MAXI/GSC 2--10 keV count rate (blue) and 2--10 keV flux measured with NuSTAR (yellow). Flux was converted to MAXI/GSC count rate using WebPIMMS. The vertical red dashed line marks MJD 60623.766, which is the time of the first scan when the source was reported to be detected by MAXI \citep{Serino2024}. Bottom: Background subtracted NuSTAR FPMA+FPMB light-curves. NuSTAR light-curves were reconstructed from both Mode 1 and Mode 06 data.  
\label{fig:lightcurves}}
\end{center}
\end{figure*}

MAXI, launched in 2009, is an all-sky X-ray monitor installed on the Japanese Experiment Module onboard the International Space Station \citep{MAXI}. It consists of two instruments: the Gas Slit Camera (GSC) with a net bandpass of 2--20\,keV and field of view (FOV) of $1.5^\circ\times160^\circ$ \citep{MAXIGSC}, and the Solid-state Slit Camera (SSC) with a net bandpass of 0.7--7\,keV and FOV of $1.5^\circ\times90^\circ$. The SSC ceased operations on June 22, 2022, and was therefore not used during observations of \source. MAXI/GSC scans the full X-ray sky once every orbit (about 92\,minutes), such that individual sources are within the FOV for about $45$\,seconds per orbit \citep{Sugizaki2011}. The resulting 3-$\sigma$ detection limit for one scan is $80$\,mCrab in the 3--10\,keV band \citep{Negoro2016}. The MAXI/GSC light-curve of \source\ available from the online MAXI light-curve portal\footnote{http://maxi.riken.jp/star\_data/J1752-457/J1752-457.html} contained data with a low signal-to-noise ratio due to contamination by the bright low-mass XRB (LMXB), H~1735$-$444, separated from \source\ by $2.8^{\circ}$ in the GSC anode direction. Therefore, we instead obtained the MAXI/GSC light-curve using the point-spread-function fit method presented by \citet{Morii2016}. We excluded scanning data if none of the cameras completely covered the source region. We show the MAXI/GSC light-curve for MJD 60622 -- MJD 60629 in the top panel of Figure \ref{fig:lightcurves}. We have re-binned the data by up to 10 orbits 
for clarity before and after the brightest phase of the burst.
We note that visual inspection of the MAXI/GSC light-curve of \source\ binned into 16 days over the course of more than 15 years reveals a hint of a weak enhancement ($\sim$ 1--2 mCrab) beginning in July, 2024, and lasting roughly through November, 2024.

\section{NuSTAR Data and Analysis}\label{sec:nustar}

NuSTAR is the first focusing hard X-ray telescope \citep{NuSTAR}. It consists of two focal plane modules, FPMA and FPMB, each of which is composed of a set of four Cadmium Zinc Telluride (CZT) detectors in a $2\times2$ configuration. FPMA and FPMB are both paired with a focusing optics module at a focal length of 10\,m. The resulting energy range of NuSTAR is 3--79\,keV, and the FOV is $10^{\prime}$ at 10\,keV.

\source\ was observed twice by NuSTAR during the tail of the 2024 superburst. The first observation was performed using a tiling strategy designed to locate the source within the MAXI/GSC 90\% confidence region. Four tiling observations were carried out with exposures of about $5$\,ks each. The source was detected in one of these observations with OBSID 81010302001. Using the resulting localization, a second, pointed observation (OBSID 81001347001) was carried out less than a day later. The start and end times and the total exposure times of the two NuSTAR observations are listed in Table \ref{tab:nustar_obs}. We extracted cleaned event lists for both observations using the standard \texttt{nupipeline} options with \textit{saamode=NONE} and \textit{tentacle=no}. We used HEASOFT v6.34, NuSTARDAS v2.1.4a, and CALDB v20241104. 

\begin{figure*}[t!]
\begin{center}
\includegraphics[width=0.44\textwidth]{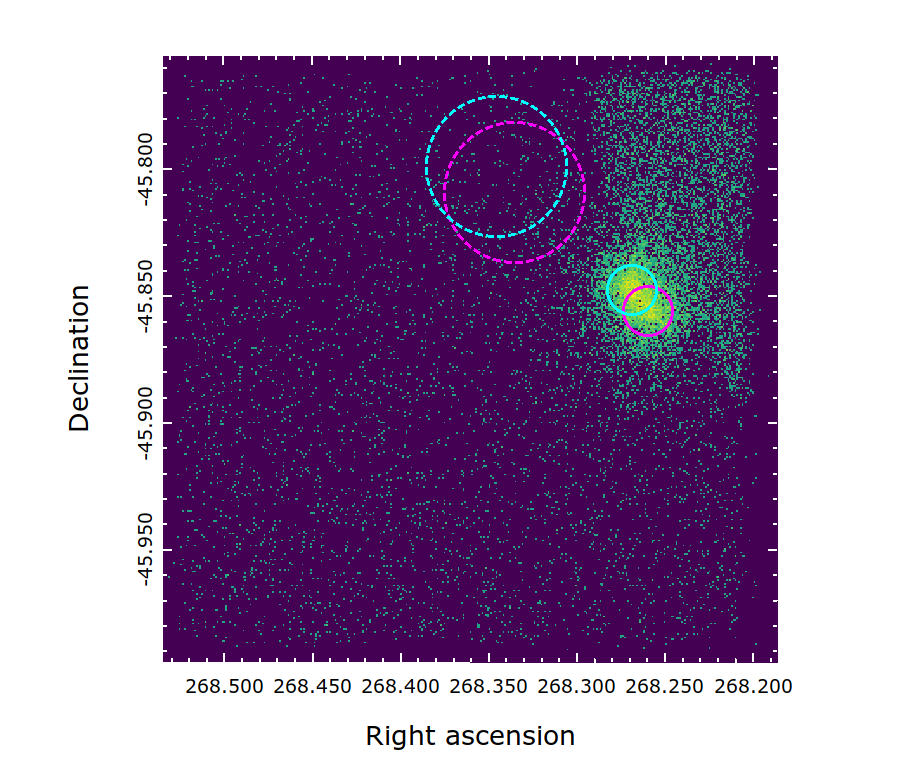}
\includegraphics[width=0.5\textwidth]{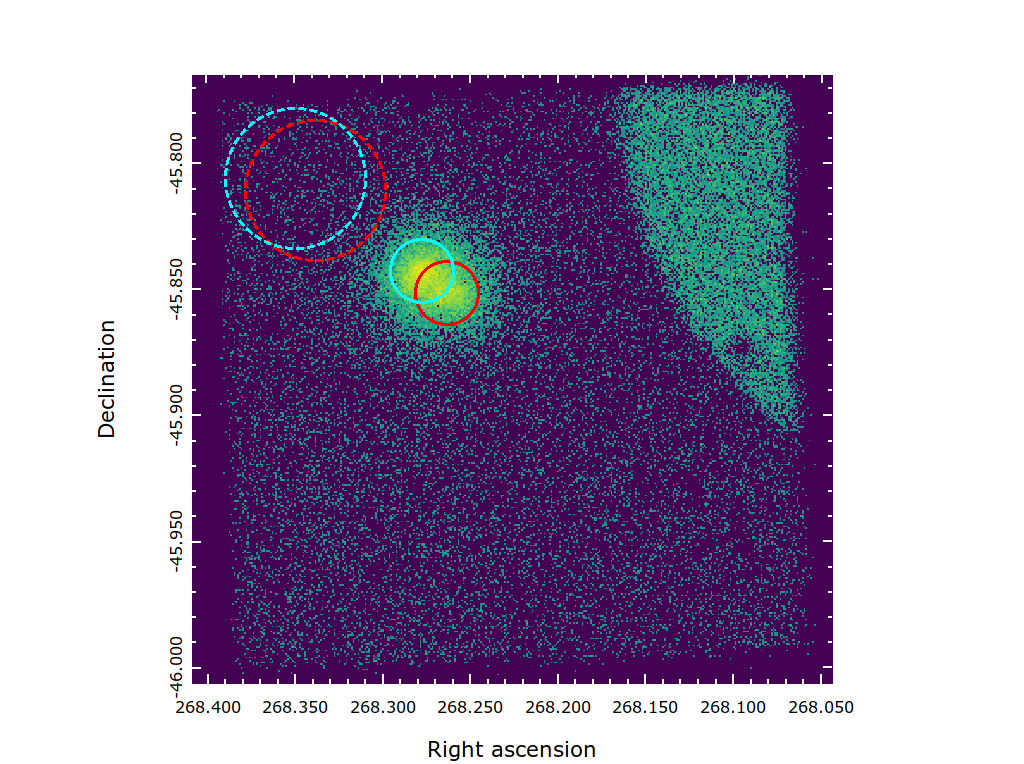}
\caption{Mode 06 images taken by FPMA during the first (left) and second (right) NuSTAR observations. We split Mode 06 data into different CHU combinations in order to mitigate the effects of centroid smearing. Source and background extraction regions are shown with solid and dashed lines, respectively, for  CHU combinations CHU23 (cyan), CHU3 (magenta), and CHU13 (red). A region of stray light can be seen in the northwest corner of each image. We attempted to choose source regions such that stray light contamination of the science products was minimized. For FPMB, the source was further away from the region of stray light during the first observation, allowing us to use slightly larger source regions than we chose for FPMA.
\label{fig:images}}
\end{center}
\end{figure*}

During both NuSTAR observations, the source was at a low SAA, leading to complications in the image reconstruction process. At low SAA, the primary star tracker, Camera Head Unit 4 (CHU4), which is located on the X-ray optics bench, may be blinded by the Sun and therefore become unusable for image reconstruction. The usual science mode, which utilizes CHU4, is called Mode 1. Outside of Mode 1, image reconstruction must be performed using combinations of CHUs 1, 2, and 3, which are located on the spacecraft bus. This is referred to as Mode 06 data. Depending on the pointing of the spacecraft any given time, one of seven combinations of these three star trackers may be used to produce an attitude solution. Due to thermal flexing of the mountings of CHUs 1, 2, and 3, the resulting solution may include inaccuracies which manifest as a ``smearing" of the source or the presence of multiple centroids in the reconstructed image. In order to most effectively extract scientific products from Mode 06 data, we used the \texttt{nusplitsc} command to split the cleaned event list according to active CHU combination. We performed \texttt{nusplitsc} two separate times, once with \textit{splitmode=NORMAL} and another time with \textit{splitmode=STRICT}. In some cases, strict splitting can help to remove flickering between CHU combinations, therefore mitigating multiple centroids. However, based on visual inspection using SAOImageDS9 \citep{DS9}, we determined that using the strict splitmode did not improve the image reconstruction compared to normal mode, and in fact strict mode only resulted in a reduction in the overall number of events available for science products. We therefore proceeded using the event files produced using \textit{splitmode=NORMAL}. 

We used DS9 to inspect the event files from Mode 1 and the split event files from Mode 06. For each file, we chose a circular source extraction region and centered it on the source using the DS9 centroid calculation feature. During the first observation, there was significant stray light which did not overlap with the source position, but which was close enough to the source to become a consideration when choosing the source region. In order to avoid stray light contamination of the source data, we chose source regions with radius $35^{\prime\prime}$ for FPMA data and $40^{\prime\prime}$ for FPMB data from the first observation. For both FPMA and FPMB data from the second observation, we chose source regions with radius $45^{\prime\prime}$. For all event files, we chose circular background regions with radius $100^{\prime\prime}$. The stray light region, characteristic centroid smearing, and examples of the source and background regions we chose for FPMA are shown in the Mode 06 images for both observations in Figure \ref{fig:images}. After selecting extraction regions, we used \texttt{nuproducts} to produce source and background spectra and light-curves for each cleaned event file. The background-subtracted light-curves (summed FPMA and FPMB) are shown in the bottom panels of Figure \ref{fig:lightcurves}. Finally, for each observation, we used \texttt{addspec} to sum the Mode 1 and Mode 06 data into one pair of FPMA source and background spectra and one pair of FPMB spectra. We used \texttt{ftgrouppha} to rebin the spectra according to the method described by \cite{Kaastra2016} with a minimum of 10 counts per bin.

\subsection{Spectral Analysis}\label{sec:spectrum}

\begin{figure*}[t!]
\begin{center}
\includegraphics[width=0.95\textwidth]{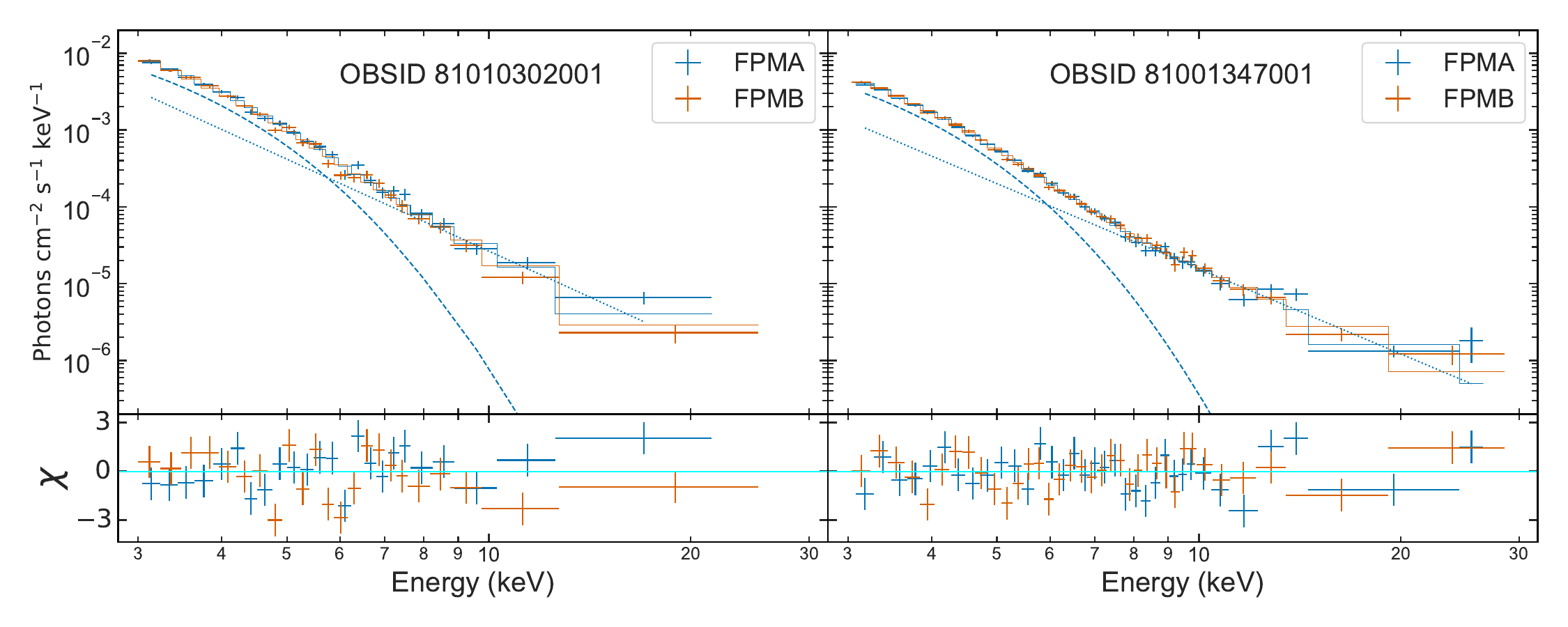}
\caption{NuSTAR FPMA (blue) and FPMB (orange) spectra of the source during the first (left) and second (right) observations. The dashed and dotted lines show the blackbody and power law components of the spectral model, respectively.
\label{fig:nustar_spectrum}}
\end{center}
\end{figure*}

\begin{deluxetable*}{cccc}
\tablenum{2}
\tablecaption{Spectral parameters and unabsorbed 2--10\,keV flux measured during the NuSTAR observations. \label{tab:spectral_params}}
\tablewidth{0pt}
\tablehead{
\colhead{Model Component} &  \colhead{Parameter}    & \colhead{Obs1 value}   & \colhead{Obs2 value}}
\startdata
\multirow{3}{*}{bbodyrad}    & $kT\,\mathrm{(keV)}$      &  $0.61\pm0.02$ & $0.60\pm0.01$  \\
                    & $R\,\mathrm{(km)}$\tablenotemark{$\dagger$}      &  $7.7^{+1.2}_{-1.4}$ & $6.0^{+0.4}_{-0.3}$  \\
                    & $f_\mathrm{2-10}\,\mathrm{(erg\,cm^{-2}\,s^{-1})}$ & $7^{+1}_{-2}\times10^{-11}$ & $4.3^{+0.3}_{-0.4}\times10^{-11}$ \\
\hline
\multirow{2}{*}{powerlaw}    & $\Gamma$       &  $4.0\pm0.4$ & $3.7\pm0.2$  \\
    & $f_\mathrm{2-10}\,\mathrm{(erg\,cm^{-2}\,s^{-1})}$ & $5^{+3}_{-2}\times10^{-11}$ & $2.1^{+0.6}_{-0.5}\times10^{-11}$ \\
\hline
\hline
    &   $\chi^2/\mathrm{d.o.f.}$    & $91/57$ & $109/91$ \\
    &   $f_\mathrm{2-10}\,\mathrm{(erg\,cm^{-2}\,s^{-1})}$ & $(1.2 \pm 0.1)\times10^{-10}$ & $6.4^{+0.3}_{-0.2} \times10^{-11}$ \\
\enddata
\tablenotetext{\dagger}{Radius calculated assuming a distance of $8$\,kpc, without correcting for scattering in the neutron star atmosphere.}
\end{deluxetable*}

We analyzed the spectra using Xspec \citep[v12.14.1][]{Xspec}. In order to perform parameter estimation, we used the $W$ fit statistic \citep{Wachter1979}, which is used by Xspec in the case of Poisson source and background spectra. We used the Chi-square test statistic to assess goodness-of-fit. For both observations, the number of background counts becomes comparable to the number of source counts above around $30$\,keV. Therefore we only modeled the spectra for energies in the range 3--30\,keV. Throughout our analysis, we performed simultaneous fits to the FPMA and FPMB spectra, applying a multiplicative constant to account for small differences in the overall normalization between the two spectra.

Due to the lack of evidence for strong absorption and to the insensitivity to absorption in the NuSTAR band, for all spectral models we applied an absorption column with fixed density using the multiplicative Xspec model component, \texttt{tbabs}. We fixed the column density to a value of $N_\mathrm{H}=10^{21}\,\mathrm{cm^{-2}}$, determined using the full-sky HI survey, HI4PI \citep{HI4PI}. We first attempted to model the NuSTAR spectra using a simple absorbed power law with a high-energy cutoff. We found that this did not produce an adequate fit, resulting in a test statistic of $\chi^2/\mathrm{d.o.f.}=142/58$ for the first observation and $\chi^2/\mathrm{d.o.f.}=286/92$ for the second observation. 

Given the superburst interpretation presented in \citet{Aoyama2025}, we added a physically-motivated single-temperature blackbody component to our spectral models, representing thermal emission from the surface of a neutron star heated by either thermonuclear burning during the burst or by persistent accretion flow. Notably, the peak flux observed by MAXI/GSC during the burst was around $2\times10^{-8}$\,\fluxcgs. Although the distance to \source\ is unknown, the source is located at $(\alpha, \delta) = (268.269, -45.866)$, or $(l,b)=(346.092,-9.8517)$, placing it at a $17^{\circ}$ separation from the Galactic center. Like \citet{Aoyama2025}, we tentatively assume that the angular proximity of the source to the Galactic center translates into physical proximity, giving a distance of around 8\,kpc. This assumption leads to a derived peak luminosity on the order of $10^{38}$\,\lumcgs, very near the Eddington luminosity for spherical accretion onto a neutron star: $L_\mathrm{Edd}=1.38\left(M/{M_\odot}\right)\times10^{38}$\,\lumcgs, where the canonical neutron star mass is $M=1.4M_\odot$. Thus, the physical interpretation of the source as a bursting neutron star is consistent with a distance of around 8\,kpc, and we therefore present derived luminosities and blackbody radii using this fiducial value.

We found that when we included a single-temperature blackbody component, the value of the high-energy cutoff was pushed far beyond the energy range we considered, indicating that the cutoff is not necessary to fit the data. The combination of a single-temperature blackbody and a simple power law component fits the data significantly better than a cutoff power law alone, resulting in a test statistic of $\chi^2/\mathrm{d.o.f.}=91/57$ for the first observation and $\chi^2/\mathrm{d.o.f.}=109/91$ for the second observation. We found that the spectra can be described without significant changes in the blackbody temperature and the spectral index between the two observations. Instead, the changes between the spectra can be almost entirely described by changes in overall flux, with both the blackbody flux (i.e. radius) and power law flux decreasing between the first and second observations. We measured a decrease in unabsorbed flux of almost one half from $(1.2\pm0.1)\times10^{-10}$\,\fluxcgs during the first observation to $(6.4^{+0.3}_{-0.2})\times10^{-11}$\,\fluxcgs during the second observation, corresponding to luminosities of around $9\times10^{35}$\,\lumcgs and $5\times10^{35}$\,\lumcgs.

In Table \ref{tab:spectral_params}, we list the resulting blackbody and power law parameters, the test statistic achieved with the model, and the unabsorbed 2--10\,keV flux measured for both observations. In Figure \ref{fig:nustar_spectrum} we show the FPMA and FPMB spectra, with the first observation shown on the left and the second observation shown on the right. We also plot the best-fit model including the single-temperature blackbody and the power law components. In addition to the good test statistic achieved with this model for the second observation, we do not observe residual features which would indicate that an additional component is necessary to describe the spectra.

We also considered slightly different forms of this basic spectral model. For example, we replaced the single-temperature blackbody with a multi-temperature disk blackbody. This model resulted in a comparable or slightly worse test statistic (e.g. $\chi^2/\mathrm{d.o.f.}=112/91$ for the second observation). This model yields an inner disk temperature of around $kT_\mathrm{in}=0.7$\,keV and a slightly shallower power law with a photon index of $\Gamma = 3.4$. Similarly, replacing the simple power law component with a thermally Comptonized continuum component \citep[\texttt{nthcomp;}][]{Zdziarski1996,Zycki1999}, did not improve the test statistic ($\chi^2/\mathrm{d.o.f.}=110/91$ for the second observation). In this case, the relevant spectral parameters did not change compared to the simple power law model. 

In order to check the self-consistency of the spectral model, we also reproduced the spectral fits using \texttt{simpl} \citep{simpl} rather than an analytic power law. We again found no appreciable change in the test statistic ($\chi^2/\mathrm{d.o.f.}=110/91$ for the second observation). This model yields a slightly larger blackbody radius (at 8\,kpc, uncorrected for atmospheric scattering effects) of $11^{+2}_{-1}$\,km and $7.4^{+0.6}_{-0.5}$\,km for the first and second NuSTAR observations, respectively. For both observations, using \texttt{simpl} instead of an analytic power law results in a decrease in the blackbody temperature by a few percent, an increase in the photon index of about 0.1, and an upscattering fraction of around $10\%$, assuming that the power law is seeded by the spherical blackbody component.

We note that the test statistic is comparably poor for the first observation, leading to a null-hypothesis probability of $0.3\%$. We find that the residuals for these spectra are not well matched between FPMA and FPMB (see the bottom left panel of Figure \ref{fig:nustar_spectrum}). We therefore hypothesize that the higher test statistic is primarily due to systematic error introduced by the complicated Mode 06 extraction procedure with source regions bordering the stray light region. While the inferred parameters provide a point of comparison with the second observation, we do not rely heavily on the first observation to inform our conclusions.

\subsection{Timing analysis}\label{sec:timing}

\begin{figure*}[t!]
\begin{center}
\includegraphics[height=0.38\textwidth]{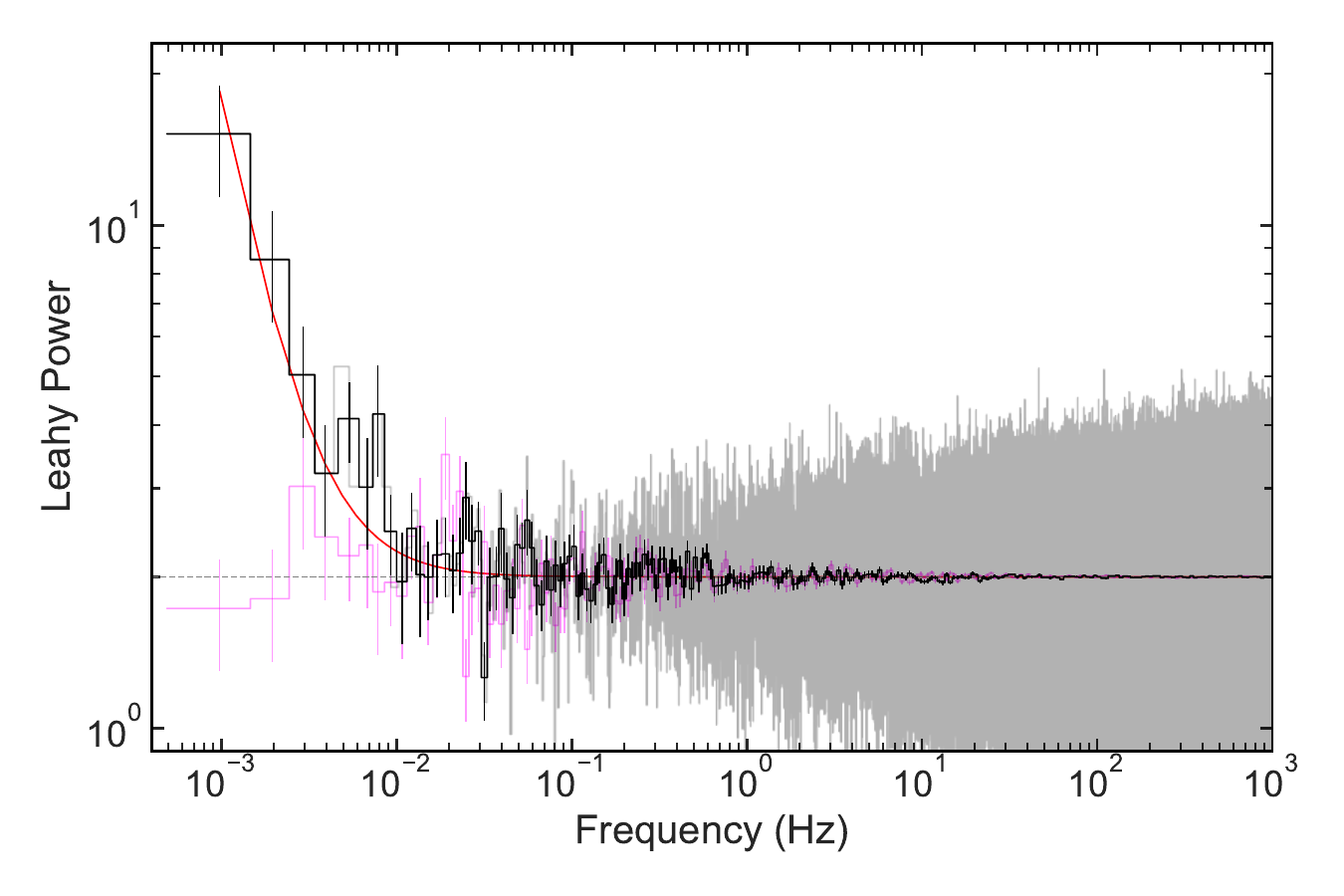}
\includegraphics[height=0.38\textwidth]{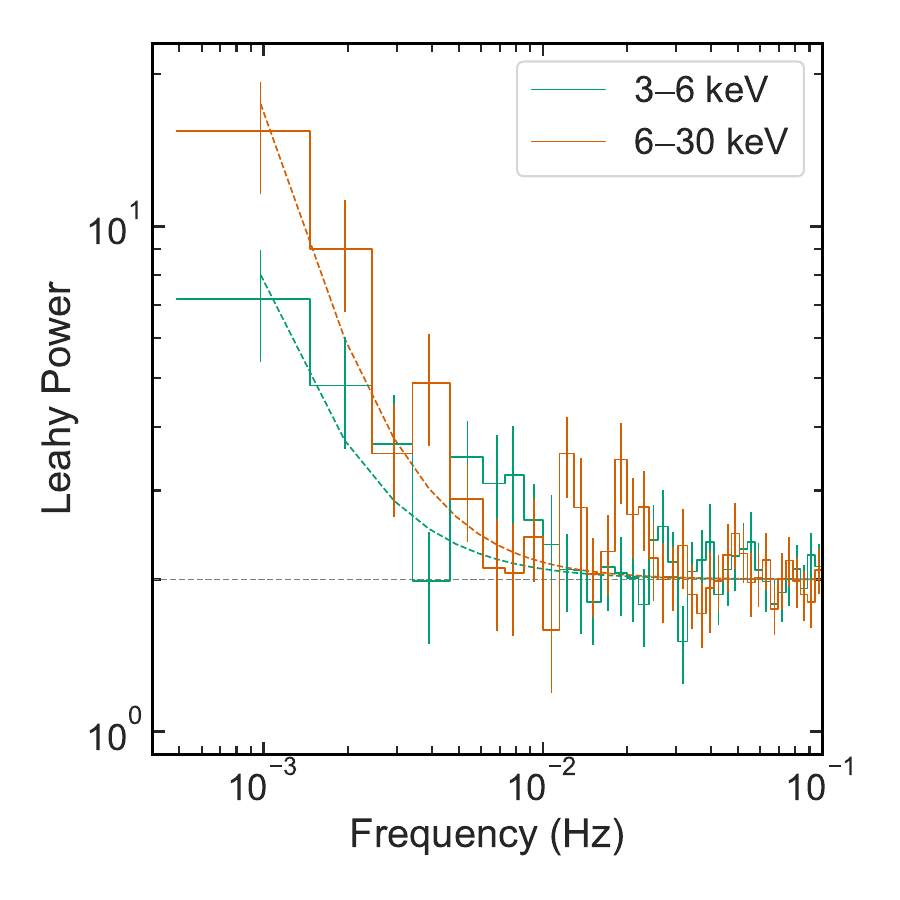}
\caption{Left: Leahy-normalized power density spectrum (PDS) of the 3--30\,keV NuSTAR source (black) and background (magenta) light-curves measured during the second observation. The PDS have been binned logarithmically for clarity, and the unbinned source PDS is shown in gray. The only clear feature we observe is a low-frequency noise component, which we fit with a power law, shown in red. Right: Leahy-normalized source PDS calculated for the 3--6\,keV (green) and 6--30\,keV (orange) light-curves. The low-frequency component is somewhat more pronounced in the hard X-ray band. The effects of deadtime were corrected using the Fourier Amplitude Difference (FAD) normalization technique.
\label{fig:pds}}
\end{center}
\end{figure*}

\begin{deluxetable}{lcc}
\tablenum{3}
\tablecaption{Details of the NuSTAR timing analysis. \label{tab:timing}}
\tablewidth{0pt}
\tablehead{
\colhead{Energy band} &  \colhead{Power law index}    & \colhead{Source rms (\%)}\\
(keV)   & & ($10^{-3}$ -- $10^{-1}$\,Hz)}
\startdata
3--30  &   $1.8^{+0.3}_{-0.2}$    &   $17\pm3$  \\
3--6   &   $1.8^{+0.7}_{-0.4}$    &   $13\pm4$  \\
6--30  &   $2.0^{+0.4}_{-0.3}$    &   $58\pm8$  
\enddata
\end{deluxetable}

We investigated the timing properties of \source\ during the second NuSTAR observation. Beginning with the clean Mode 1 and Mode 06 event files, we first corrected the photon arrival times for the motion of NuSTAR about the barycenter of the Solar System using the \texttt{ftools} utility, \texttt{barycorr}. We specified the source position as measured by Swift/XRT and reported by \cite{Liu2024}. We used NuSTAR clockfile 20100101v203, produced on April 15, 2025. Next, we filtered the events using circular source regions with radii $100^{\prime\prime}$ in order to encompass as many source counts as possible given the centroid smearing in Mode 06. Unlike our spectral analysis, and in order to avoid complicated good time intervals (GTIs), we did not split the Mode 06 events into different CHU combinations.

We next used the timing analysis software packages HENDRICS and Stingray \citep{Stingray1,Stingray2,Stingray3} to complete the rest of the timing analysis. We found that the background dominates the source spectrum above 30\,keV, so we filtered out events with energy below 3\,keV and above 30\,keV. For each focal plane module, we merged the Mode 1 and Mode 06 event lists. Using Stingray, we converted the event lists into light-curves with bins of $dt=2048^{-1}$\,s. We split these light-curves into 16 segments of length $1024$\,s each. We calculated the averaged power density spectrum (PDS) over these segments for FPMA and FPMB, then produced a combined PDS which we corrected for deadtime using the Fourier Amplitude Difference normalization described in \cite{FAD}. We repeated this procedure in order to produce a background PDS, using only events in background regions with radius $150^{\prime\prime}$, away from the source. The resulting Leahy-normalized PDS, binned logarithmically for clarity, are shown in the left panel of Figure \ref{fig:pds}, where the source PDS is shown in black, and the background PDS is shown in magenta. In order to investigate the energy-dependence of the source variability, we also calculated the source PDS for events with 3--6\,keV and for events with 6--30\,keV, shown in green and orange, respectively, in the right panel of Figure \ref{fig:pds}.

The PDS is relatively featureless aside from a clear low-frequency noise component. We do not find any evidence for pulsations or quasiperiodic oscillations (QPOs). We performed a least-squares fit of the PDS with a power law model. We also fit this model to the ``soft" (3--6\,keV) and ``hard" (6--30\,keV) PDS. The best-fit power law index for each energy band is listed in Table \ref{tab:timing}, along with the fractional rms calculated for the frequency range $10^{-3}$ -- $10^{-1}$\,Hz. Our results indicate that the low-frequency noise may be slightly steeper in the hard band, but the power law index is consistent with 2 within the $90\%$ confidence intervals regardless of the energy band: for the 3--6\,keV range, we find a power law index of $1.8^{+0.3}_{-0.2}$, and for the 6--30\,keV range, we find a power law index of $2.0^{+0.4}_{-0.3}$. Notably, the low frequency noise is significantly higher in the hard band than in the soft band.

\section{Discussion}\label{sec:discussion}

\subsection{Revisiting the power law index}\label{sec:gamma_discussion}

MAXI and NinjaSat observations of \source\ during a superburst in November, 2024, showed that the flux and spectral evolution were consistent with an accretion-induced superburst \citep{Aoyama2025}. While these observations were sensitive to the blackbody component of the spectrum, and have demonstrated clear evolution of the temperature and radius of the blackbody, NuSTAR's unique sensitivity above $10$\,keV has allowed us to put far better constraints on the blackbody parameters following the rapid decay of the burst. We found that by the time of the second NuSTAR observation, which occured about $79$\,hours after the onset of the burst, the blackbody temperature had decreased to $0.60\pm0.1$\,keV, and the inferred radius (assuming a distance of $8$\,kpc) had shrunk to $6.0^{+0.4}_{-0.3}$\,km. 

\begin{figure}[t!]
\begin{center}
\includegraphics[width=0.48\textwidth]{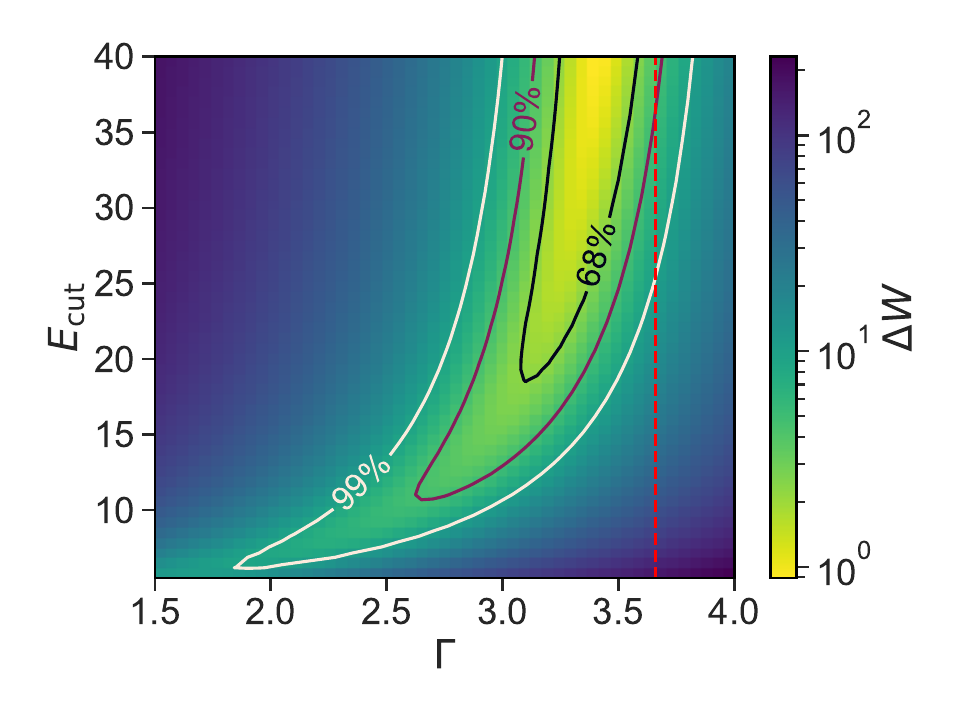}
\caption{The change in $W$-statistic with respect to the best-fit value when modeling the spectra from the second NuSTAR observation with an absorbed blackbody and cutoff power law. $\Delta W$ is plotted as a function of the power law photon index, $\Gamma$, and cutoff energy, $E_\mathrm{cut}$. The $68\%$ (1-$\sigma$), $90\%$, and $99\%$ confidence regions for these parameters are shown as black, purple, and white contours, respectively. The dashed red line marks the best-fit value, $\Gamma=3.7$, when there is no high-energy cutoff. When the cutoff is included, $E_\mathrm{cut}$ is unconstrained, with a $99\%$ lower limit of 8\,keV.
\label{fig:contour}}
\end{center}
\end{figure}

NuSTAR observations also revealed that, in addition to a blackbody component, the source spectrum contained a hard X-ray excess which is well-described by a steep power law with photon index $\Gamma\approx4$. This indicates the presence of a non-thermal plasma in which soft X-ray photons gain energy via Compton upscattering. Accreting neutron stars at similar luminosities ($10^{35-36}$\,\lumcgs) tend to exhibit power law spectral components with photon indices in the range $1.5<\Gamma<3.0$ \citep{Wijnands2015}. We therefore revisit the spectra obtained during the second NuSTAR observation in order to explore the possibility of a more typical photon index. 

As we noted in Section \ref{sec:spectrum}, when we allowed for a high-energy cutoff to be applied to the power law component, the best-fit value of $E_\mathrm{cut}$ was far beyond the NuSTAR energy range, indicating that a power law without a cutoff energy is preferred. However, including a high-energy cutoff allows for more flexibility in the photon index parameter as a low cutoff energy can compensate for a shallower power law. For the cutoff power law model, the $90\%$ confidence interval for the photon index is expanded, giving $\Gamma=3.7^{+0.2}_{-0.7}$. Furthermore, we calculated the $99\%$ confidence intervals for the photon index and cutoff energy by varying each parameter independently until the fit statistic had increased by 6.635. We found intervals of $2.3<\Gamma<4.0$ and $E_\mathrm{cut}>8$\,keV. In order to explore the parameter space in even more detail, we stepped over the photon index and the cutoff energy and calculated the $W$-statistic at each step. The resulting grid is shown in Figure \ref{fig:contour}, with contours representing the 1-$\sigma$, $90\%$, and $99\%$ confidence regions. We again find that the spectra strongly prefer a steep power law with a $90\%$ confidence region that indicates a photon index greater than 2.5 and a cutoff energy above 10\,keV. While photon index values of $<2.5$ are allowed within the 99\% confidence region, they would require cutoff energies of less than 10\,keV. This combination of parameters would imply a high optical depth, $\tau>7$ \citep[][Eq. 34]{Kara2025}.
While we cannot completely rule out a more typical value for the photon index, spectral modeling of \source\ strongly favors an abnormally steep power law in order account for the hard X-ray excess above about 7\,keV. 

We note that although \citet{Aoyama2025} reported a similarly steep power law during this phase of the outburst decay, our measurement represents a distinct result for the following reasons. First, NinjaSat is sensitive to the 2-10\,keV emission, whereas the observations presented here provide hard X-ray sensitivity. This is demonstrated clearly by our ability to disentangle multiple spectral components. To this point, \citet{Aoyama2025} found a steep photon index when modeling the source spectra using a power-law-only model. As we have shown, the soft X-ray is dominated by blackbody emission up to around 6\,keV. Therefore the evolution of the photon index presented by \citet{Aoyama2025} should be interpreted as the softening of the blackbody component rather than a measurement of the power law component itself. Finally, where the NinjaSat spectra at this phase of the decay were limited in their constraining power --- \citet{Aoyama2025} report three degrees of freedom for the fit with the steepest photon index --- the observations we have presented allowed for tighter constraints and a detailed investigation of the parameter space as we have described above.

\subsection{Prior X-ray studies of superbursts}

Existing analyses of the hard X-ray spectra of accreting neutron stars during and after superburst are limited due to the relative infrequency of these bursts. In fact, at the time of writing, we are unable to identify any published observations of superbursts or superburst decays with NuSTAR. However, the Rossi X-ray Timing Explorer (RXTE), thanks to its All-Sky Monitor (ASM) instrument, observed 12 superbursts in the hard X-ray during its lifetime \citep{Zand2017}. Often, analyses of the spectral evolution of superbursts with RXTE fixed the power law component to the value measured for persistent emission, typically around $\Gamma\approx2$ \citep[e.g.][]{Keek2008}. In some instances, significant softening of the superburst decay spectrum compared to the persistent spectrum has been observed \citep[e.g.][]{Kuulkers2004}. 

Studies of superbursts in the soft X-ray band ($<10$\,keV), although insensitive to the hard X-ray emission, have hinted at high photon indices during or shortly after superbursts or similar bursts. For example, \cite{Bult2021} reported on NICER measurements of the spectral evolution of IGR~J17062$-$6143 during an intermediate duration Type I X-ray burst observed in 2020. While not categorized as a superburst, the decay of this burst lasted several days, making it comparable to the case of \source. These authors reported that, when a relativistic reflection component was included in the spectral model of the decay emission, the photon index reached values in the range of $3<\Gamma<5$ throughout the two-day period following the burst before settling to more typical values around $\Gamma\approx2.5$. Notably, however, when the spectrum was modeled without relativistic reflection, the opposite trend in the power law photon index was recovered, with $\Gamma$ increasing from around 1 to 2 throughout the burst decay. 

In a recently published NICER study of the 2021 superburst of 4U~1820$-$30, \cite{Peng2025} modeled the persistent, burst, and decay emission using a combination of a spherical blackbody and a Comptonization component represented by the Xspec model \texttt{compTT}. The authors noted that during the burst, the contribution of the Comptonization component could be neglected, but they reported that during the first few hours of the decay, they measured a significantly lower electron temperature than during the persistent emission, indicating a softened hard X-ray component. \citet{Jaisawal2025} presented a separate study of the same observations. These authors modeled the emission using a combination of two blackbody components and an absorbed power law. Similarly to \citet{Peng2025}, these authors reported that the power law index decreased from around 2.8 back to a stable value of around 1.6 within six hours of the onset of the burst.

These NICER studies provide evidence for significant changes to non-thermal emission from accreting neutron stars during and after superbursts, but they also highlight the difficulty of disentangling spectral components with soft X-ray observations alone when the emission is dominated by a bright blackbody component. NuSTAR, on the other hand, is uniquely able to probe the hard X-ray component directly, making it possible to overcome degeneracies between the power law slope and blackbody parameters. Our result is, to the best of our knowledge, the first time that the the power law component has been constrained with hard X-ray observations in the aftermath of a superburst. The observations and results we have presented here differ from the results discussed above in other important ways. First, we have measured a steep power law index more than 5 days after the onset of the superburst in \source. In contrast, \citet{Bult2021} showed that the power law index returned to more typical values within around two days after the intermediate duration Type I X-ray burst exhibited by IGR~J17062$-$6143 in 2020, and both \citet{Peng2025} and \citet{Jaisawal2025} showed that the hard X-ray component in the spectra of 4U~1820$-$30 returned to a steady state within only a few hours of the 2021 superburst. Furthermore, 4U~1820$-$30 and IGR~J17062$-$6143 are persistently accreting or long-term transient sources, whereas \source\ appears to be a transient accretor.

\subsection{Hard X-ray variability and disk-corona coupling}

Our analysis of the source variability did not reveal discrete features like pulsations or quasiperiodic oscillations (QPOs). Instead, we found that the source variability was dominated by red noise, which we modeled with a power law. We found a power law index of $1.8^{+0.3}_{-0.2}$, which is typical of the low-frequency noise seen in the low-luminosity state of X-ray bursters \citep{Belloni2002}, and we found that the low-frequency variability is more pronounced at higher energies. Using the Xspec tool, \texttt{cpflux}, we determined that the spectral power law component makes up $22\%$ of the source photon flux in the 3--6\,keV range and $77\%$ of the source photon flux in the 6--30\,keV range. Given a total source rms of $(13\pm4)\%$ and $(58\pm8)\%$ in the soft and hard bands, respectively, and if we assume that the non-thermal component is the sole source of red noise (i.e., the rms variability of the blackbody component is very small), we infer that the $10^{-3}-10^{-1}$\,Hz rms of the non-thermal component was $(59\pm18)\%$ in the 3--6\,keV range and $(75\pm10)\%$ in the 6--30\,keV range. In other words, we find that the energy-dependent photon flux fraction of the spectral power law component is consistent with our measurements of the source rms as a function of energy under the assumption that the non-thermal component is the primary source of the low-frequency variability. 

The significant difference in the degree of variability between the thermal and power law components raises the question of whether the source remained in the radiative cooling superburst decay phase during the NuSTAR observations, or whether the source had returned to an accretion-dominated flux state.
Several pieces of evidence point towards accretion-powered emission. First, despite the fact that the two NuSTAR observations took place nearly a day apart, we do not observe evidence for cooling of the blackbody component in this interval, suggesting that the blackbody temperature had reached a persistent value. If we assume a constant blackbody radius between the first and second NuSTAR observations, we would expect to find a temperature of $kT\approx0.53$\,keV during the second observation, given the decrease in flux. We attempted to fit the spectrum observed during the second observation while fixing the blackbody radius to the value measured for the first observation, but we found that this model could not fit the data, resulting in a test statistic of $\chi^2/\mathrm{d.o.f.}=174/92$ (compared to $\chi^2/\mathrm{d.o.f.}=109/91$ when the radius was allowed to vary).

We also find that the flux ratio between the blackbody component and the power law component did not change significantly between the two NuSTAR observations, suggesting that both the thermal and power law components share a common source of energy in the form of an accretion disk. This interpretation is supported by the results of our timing analysis.
Low-frequency variability is typically attributed to variable thermal disk emission for both black hole and neutron star sources \citep{Titarchuk2007}. Given that the low-frequency noise we observe can be attributed primarily to the power law spectral component, rather than to the thermal component, this would suggest that the corona is seeded by soft X-ray photons from an accretion disk such that disk variability propagates through the Comptonizing medium. Importantly, however, the flux we measured with NuSTAR remains above the pre-superburst flux. 
We retrieved Swift XRT observations of the source in the weeks leading up to the superburst and found that the source flux (0.3-10\,keV) varied between $5\times10^{-12}$\,\fluxcgs and $5\times10^{-11}$\,\fluxcgs.
On October 24, 2024, about 16 days prior to the onset of the superburst, EP measured a 0.3-10\,keV flux of $(1.3\pm0.3)\times10^{-11}$\,\fluxcgs. We also retrieved spectral data products from Swift XRT observations taken on the same day (OBSID 00016764023) and performed a simple analysis. We found that the spectrum could be described by a power law model alone ($\chi^2/\mathrm{d.o.f.}=192/161$), or by the combination of a spherical blackbody and a power law component ($\chi^2/\mathrm{d.o.f.}=183/159$). The resulting blackbody temperature and radius are $kT=0.37\pm0.6$\,keV and $R=5\pm2$\,km (assuming a distance of 8\,kpc). Notably, the photon index for the combined blackbody and power law model is $\Gamma=1.1^{+1.2}_{-0.1}$. This value is more typical of accreting neutron stars than the steep photon index we observed in the NuSTAR spectra.

During the second NuSTAR observation, we measured a source flux (2-10\,keV) of $6.4^{+0.3}_{-0.2}\times10^{-11}$\,\fluxcgs. The flux at this time was therefore still a factor of several times higher than the per-superburst flux. The drop in flux that we observed between the first and second NuSTAR observations is also consistent with the continued downward trend in flux observed by NinjaSat at the same time. Additionally, the pre-superburst XRT spectrum was consistent with the blackbody and power law model we used to describe the NuSTAR spectra. If we assume that accretion powers the flux in both cases, then the lower flux, lower blackbody temperature, and smaller blackbody radius we measured for the XRT spectrum is consistent with a lower rate of accretion. Importantly, however, we cannot claim that the spectra strongly prefer this model over the simple power law model.

\section{Conclusions}\label{sec:conclusions}

We have presented an analysis of two NuSTAR observations of \source\ following a superburst which occurred in November, 2024. We constrained the blackbody temperature and radius, which describe the thermal component of the X-ray spectrum, and we did not find evidence for significant cooling in the roughly 24 hours between the first and second NuSTAR observations. Timing analysis of the NuSTAR light-curve suggested that the non-thermal spectral component was seeded by photons from an accretion disk rather than from the neutron star surface. We therefore conclude that at the time of the NuSTAR observations, the source had entered a flux state powered by disk accretion rather than radiative cooling.

We also showed that the NuSTAR spectrum exhibited a steep power law tail with photon index $\Gamma\approx4$ between 4 and 6 days after the onset of the burst, representing the first time that the power law component has been constrained with hard X-ray observations in the aftermath of a superburst. Accreting neutron stars typically exhibit power law photon indices in the range $1.5<\Gamma<3$, making our measurement of $\Gamma\approx4$ anomalously steep. In the context of the aftermath of a superburst, this result may point towards a unique coronal state due to interaction with the burst emission. However, our ability to interpret our results in this context is limited by the lack of NuSTAR observations prior to the superburst and throughout the superburst decay.
Our results demonstrate the importance of rapid follow-up and high cadence observations of superbursts with focusing hard X-ray observatories like NuSTAR in order to place superior constraints on the spectral evolution and physical properties of bursting sources. Future hard X-ray observations of superbursts will help to elucidate the behavior of the corona before, during, and after these long-lasting thermonuclear flashes.

\begin{acknowledgments}
This work was partially supported under NASA Grant No. 80NSSC25K7455 and made use of data from the NuSTAR mission, a project led by the California Institute of Technology, managed by the Jet Propulsion Laboratory, and funded by the National Aeronautics and Space Administration.
This work was also supported by Grants-in-Aid for Scientific Research 21K03620 (H.N.) from the Ministry of Education, Culture, Sports, Science and Technology (MEXT) of Japan.
We thank the NuSTAR Operations, Software, and Calibration teams for support with the execution and analysis of these observations. This research has made use of the NuSTAR Data Analysis Software (NuSTARDAS), jointly developed by the ASI Science Data Center (ASDC, Italy) and the California Institute of Technology (USA). 
This research has also made use of the MAXI data provided by RIKEN, JAXA and the MAXI team.
We would like to acknowledge the Einstein Probe team, with particular thanks to Weimin Yuan and Dongyue Li, for their assistance in determining the pre-superburst flux of the source.
Finally, we would like to thank the anonymous reviewer for their feedback which improved the quality of this paper.
\end{acknowledgments}

\vspace{5mm}
\facilities{MAXI/GSC, NuSTAR}

\software{astropy \citep{astropy1,astropy2,astropy3},  
         Stingray \citep{Stingray1,Stingray2,Stingray3}, 
         Xspec \citep{Xspec},
         SAOImageDS9 \citep{DS9}
         }

\bibliography{main}{}
\bibliographystyle{aasjournal}



\end{document}